# A Two-Stage Reconstruction of Microstructures with Arbitrarily Shaped Inclusions

Ryszard Piasecki[1*], Wiesław Olchawa[1], Daniel Frączek[2], Agnieszka Bartecka[1]

[1] Institute of Physics, University of Opole, Oleska 48, 45-052 Opole, Poland; wolch@uni.opole.pl (W.O.); jazgara@uni.opole.pl (A.B.)

[2] Department of Materials Physics, Opole University of Technology, Katowicka 48, 45-061 Opole, Poland; dfracz@gmail.com

* Correspondence: piaser@uni.opole.pl

**Abstract:** The main goal of our research is to develop an effective method with a wide range of applications for the statistical reconstruction of heterogeneous microstructures with compact inclusions of any shape, such as highly irregular grains. The devised approach uses multi-scale extended entropic descriptors (ED) that quantify the degree of spatial non-uniformity of configurations of finite-sized objects. This technique is an innovative development of previously elaborated entropy methods for statistical reconstruction. Here, we discuss the two-dimensional case, but this method can be generalized into three dimensions. At the first stage, the developed procedure creates a set of black synthetic clusters that serve as surrogate inclusions. The clusters have the same individual areas and interfaces as their target counterparts, but *random shapes*. Then, from a given number of easy-to-generate synthetic cluster configurations, we choose the one with the lowest value of the cost function defined by us using extended ED. At the second stage, we make a significant change in the standard technique of simulated annealing (SA). Instead of swapping pixels of different phases, we randomly move each of the selected synthetic clusters. To demonstrate the accuracy of the method, we reconstruct and analyze two-phase microstructures with irregular inclusions of silica in rubber matrix as well as stones in cement paste. The results show that the two-stage reconstruction (TSR) method provides convincing realizations for these complex microstructures. The advantages of TSR include the ease of obtaining synthetic microstructures, very low computational costs, and satisfactory mapping in the statistical context of inclusion shapes. Finally, its simplicity should greatly facilitate independent applications.

**Keywords:** two-stage reconstruction; multi-scale entropic descriptors; simulated annealing for clusters; random heterogeneous materials

## 1. Introduction

The modeling of microstructures is undoubtedly an important area of research in materials science. This kind of synthetic microstructures can be obtained using theoretical models or statistical reconstruction based on real data. The latter usually contain incomplete microstructural information. "An effective reconstruction procedure enables one to generate accurate structures at will, and subsequent analysis can be performed to obtain desired macroscopic properties of the media". This sentence in Torquato's book [1], briefly describes why the modeling of heterogeneous materials is so important in the computational materials science [2–8]. Even prototype three-dimensional heterogeneous structures reconstructed approximately using limited statistical information coming from the cross-section of a given sample are still an important element in modeling and predicting physical properties [9–16]. On the other hand, the advanced high-resolution imaging techniques such as transmission electron microscopy or tomographic analysis, expensive though they are regarding their applications, significantly support the modeling of the so-called structure–property relationships [17–19].

Usually, different methods of microstructure characterization and reconstruction reveal certain differences in microstructural features sensitive to length scale, cf. reconstructed speckle patterns in [20,21]. This is important from the point of view of the macroscopic properties of the material, because some of them may be sensitive to subtle microstructural differences. Hence there is a need to develop specialized methods of statistical reconstruction.

Among the various methods of reconstruction can be mentioned characteristic approaches, e.g., using support vector machines [22], based on the genetic algorithm (GA) compared with the simulated annealing (SA) and with the maximum entropy (MaxEnt) technique [23], statistical entropic descriptors (ED) [24–27], composition, dispersion, and geometry descriptors [28], two-point correlation functions and cellular automaton [29], multi-point statistics [30,31], texture synthesis [32], watershed transform and cross-correlation function [33], supervised, generative, transfer or deep learning [34–38], a shape library containing morphologies of heterogeneities extracted from micro-computed tomography images [39], a morphological completeness analysis [40], a successive calculation of conditional probability for multi-phase materials with any level of complexity [41], the Laguerre tessellation for ceramic foams preferably with not spherical cells [42] or using a single SEM foam image and a hybrid algorithm to the pore-sphere packing problem [43], to name just a few of them.

An important element in statistical reconstruction is the quantitative characterization of the associated multi-scale spatial inhomogeneity. To extract this kind of structural information, different types of descriptors can be used. Some statistical descriptors naturally define the "energy" cost function for the SA method preferred here [44–46]. For structurally complex random composites, a hybrid combination of two-point correlation function and cluster function [47] is particularly useful. On the other hand, also the hybrid combination of entropic measures of spatial inhomogeneity and statistical complexity [24,48–50] in the case of labyrinth patterns [51] and dispersed irregular islands [52] has been successfully tested. However, as a general technique, the SA approach is rather slow.

In the current attempt, we consider a new approach to materials that are structurally similar to those analyzed in [52]. However, the process of statistical reconstruction has been optimized in a completely new way. In particular, we focus on two-phase materials containing the "black-phase" represented by well-isolated compact inclusions, e.g., grains, granules, particles, or compact clusters of different sizes and any shapes that are dispersed in the "white-phase" matrix treated as background. Usually for this type of polydispersity, it is difficult to obtain synthetic microstructures with similar macroscopic properties due to the very variable spatial inhomogeneity, even on large length scales. In addition, such reconstructions require a further reduction in computational costs. In this work, we try to solve both problems as follows.

The SA standard approach poses difficulties related to the optimization of various spatial features of a microstructure during its reconstructing, e.g., when modifying clusters shape and their relative locations. Therefore, the division of these processes into separate stages, which are algorithmically easier, should significantly simplify and accelerate the entire process of statistical reconstruction. Interestingly, in relation to cluster microstructures [47] or dense disordered packings of hard convex lens-shaped particles [53], the authors used the idea of separating features directly related to the cluster properties and independent of their location by means of a two-point cluster function $C_2$ from the features of the spatial distribution of clusters characterized by a two-point correlation function $S_2$. However, these partially complementary statistical features were simultaneously calculated as part of multi-component objective function. In this spirit, focusing on the further development of the methodology of entropic descriptors within SA, we go one-step further. Namely, we separate the process of creating synthetic clusters (equivalents of target inclusions) from the process of optimizing their spatial distribution. We call these steps the first and second stages, and the proposed approach is called two-stage reconstruction (TSR).

As we have learnt recently, a somewhat similar idea of decoupling was considered by Yang et al. [39]. In the first step, the particles taken from the earlier prepared shape library were virtually packed inside the domain to build a raw periodic model of the microstructure. Then, the final virtual microstructure was created by using genetic algorithm, eliminating some of the inclusions or by simulation of their sequential relocations within the raw microstructure.

In our approach, we first prepare a set of synthetic clusters whose number and individual interfaces are consistent with the target. For each of the synthetic clusters created, the procedure begins with the initial quasi-rectangle of the required area. Then, the designed random procedure optimizes the specific two-component function to modify the interface and shape of the generated cluster.

At the second stage, instead of accidentally placing individual black pixels, we randomly generate the several trial configurations but from the previously created synthetic clusters. From these configurations, we choose the one with the lowest initial energy value for the cost function defined by the extended ED for spatial inhomogeneity. Now, as part of the Monte Carlo approach, instead of single-phase pixels, previously prepared synthetic clusters enter the game. For this purpose, we use a special SA program tailored directly to any compact clusters.

During the TSR procedure, each randomly selected synthetic cluster can move in a random direction and with a random step length if its ending position does not coincide with other clusters. Each attempt is accepted or rejected according to standard SA conditions. However, these conditions apply directly to synthetic clusters, not pixels. The movements of entire synthetic clusters are key to speeding up the statistical reconstruction process. Our findings show that this type of optimization also has an additional advantage. The current TSR produces results of similar accuracy compared to the results obtained in the standard way by the combination of two-point correlation function $S_2$ and cluster function $C_2$. However, the standard method is much less computationally efficient because it uses single pixels by definition. The examples of reconstructed sufficiently complex microstructures confirm the reliability of the TSR method using the linear path function for its validation.

The article has the following structure. First, we will introduce a specific procedure that creates a set of synthetic clusters with appropriate statistical properties. Thus, any of their random arrangements can be used as the initial configuration. Then, we briefly describe the appropriate set of three entropy descriptors. These descriptors are involved in defining the objective function for discrete length scales associated with a given target pattern. Then, we test our approach on morphologically complicated examples of silica dispersed in a rubber matrix as well as stones in a cement paste and, finally, we summarize the main results.

## 2. Creating Synthetic Clusters–Stage One

In this section, we briefly describe the first stage of our approach. In the starting point, we have a digitized binary image (target pattern) with black compact inclusions in a white matrix. Pixels are treated as 1 × 1 uniform square elements, black or white. To create a set of statistically equivalent synthetic clusters for all compact inclusions in a target pattern, we use two auxiliary functions, $f_1$ and $f_2$. For each of the compact inclusions, we minimize the difference of the respective interfaces using the simple function

$$f_1 = \left(1 - I/I_{\text{target}}\right)^2 . \tag{1}$$

Here, $I_{\text{target}}$ and $I$ are the values of the interfaces of the target inclusion and its current synthetic counterpart, respectively. By the cluster interface, we mean the sum of the "outer" sides of all edge pixels. For instance, when a 1 × 1 edge pixel has only one common side with the rest of a given cluster, the local sum of the unit "outer" sides is three. At the same time, to achieve a certain degree of similarity in shape between the considered target inclusion and related synthetic cluster, we use the second function

$$f_2 = \frac{1}{N} \sum_{i=0}^{N-1} \left[ h_{\text{target}}(i) - h(i) \right]^2 \Big/ \left( \max h_{\text{target}} \right)^2 . \tag{2}$$

This function is defined as the sum of squares of normalized differences between the values $h_{\text{target}}(i)$ and $h(i)$ of respective surface histograms. The values $h_{\text{target}}(i)$ and $h(i)$ denote the sizes of the collections of distances between the edge pixels of the considered target inclusion and related synthetic cluster. Notice that index $i$ refers to each of the groups of linear distances between the surface pixels, which belong to the subsequent intervals, $[i, i+1)$. Each of the $N$ intervals contains the respective counts, $h_{\text{target}}(i)$ and $h(i)$, for the related distances. For clarity, the case with $i=4$ for the selected target inclusion (with $h_{\text{target}}(4)=95$ such connections) and the associated synthetic cluster (with $h(4)=87$ such connections after finishing the creation procedure summarized in Table 1) is shown in Figure 2 below.

Keeping those functions in mind, the following practical approach is preferred. Suppose that for a given target inclusion, its $A_{\text{target}}$ area (in black pixels) fulfils the inequality: $n^2 < A_{\text{target}} \leq (n+1)^2$, where $n \geq 0$ is the corresponding integer. Then, the initial synthetic cluster in the form of a quasi-rectangle is created by pre-filling the square with a linear size $(n+1)$ with the $A_{\text{target}}$ number of black pixels being unit elements of the main phase. Individual columns of the square are filled from the left to the right until the $A_{\text{target}}$ capacity is exhausted.

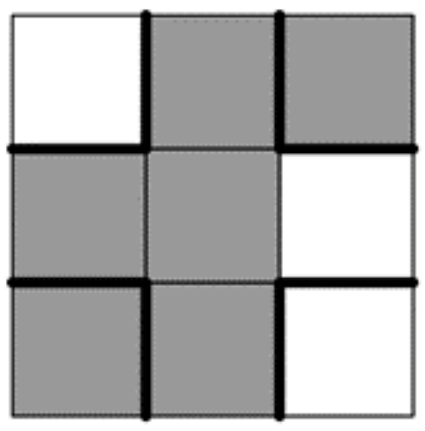

**Figure 1.** Sketch explaining the meaning of the auxiliary parameter $W_{\text{centr-black}}$ for a case where its value equals 4.

Now, the considered synthetic cluster in the form of a quasi-rectangle is modified according to some rules. To find the so-called central black pixel, we randomly select a black pixel belonging to the surface of the non-trivial synthetic cluster until the parameter $W_{\text{centr-black}} = 2$. This value describes the needed number of walls among those marked by bold lines in Figure 1 that separate the black pixels from the white pixels currently located in the Moore neighborhood. For the case shown in Figure 1, according to the definition $W_{\text{centr-black}} = 4$, the random selection of the central black pixel must be repeated. Similarly, the central white pixel belonging to the surroundings of the considered synthetic cluster is randomly selected until $W_{\text{centr-white}} = 2$. After a successful selection of the central black pixel and central white pixel, the central pixels are swapped in this pair. Accordingly, the two above conditions exclude the risk of dividing the synthetic cluster into parts, as well as the appearance of pores within this cluster. Let $Q$ be the current number of rejected attempts. We define $Q_{\max} \equiv 3 \times$ (the number of surface pixels of the considered cluster) as the maximum number of subsequent rejections. Table 1 shows all possible cases during the process of creating a synthetic cluster.

**Table 1.** Summary activities for all possible situations when creating a synthetic cluster.

| Actions | $f_1(new) < f_1(old)$ | $f_2(new) < f_2(old)$ | $Q > Q_{\max}$ |
|---|---|---|---|
| accept attempt and set $Q \leftarrow 0$ | 1 | 1 | * |
| | 1 | 0 | 1 |
| reject attempt and increase $Q \leftarrow Q + 1$ | 1 | 0 | 0 |
| | 0 | * | * |

1 = true, 0 = false, * ∈{true, false}.

Summarizing, the function $f_1$ never increases, while the function $f_2$ can do so, but after a long series of unsuccessful attempts. The optimization of both goal functions is stopped when the arbitrarily set total number of attempts exceeds $3 \times 10^3 \times Q_{\max}$.

In Figure 2, we present the edges of the chosen exemplary target inclusion and its synthetic counterpart that was prepared using the above way. In addition, for the chosen length interval $[i = 4, i+1 = 5)$, the corresponding collections $h_{\text{target}}(i)$ and $h(i)$ are displayed for the linear distances between edge pixels.

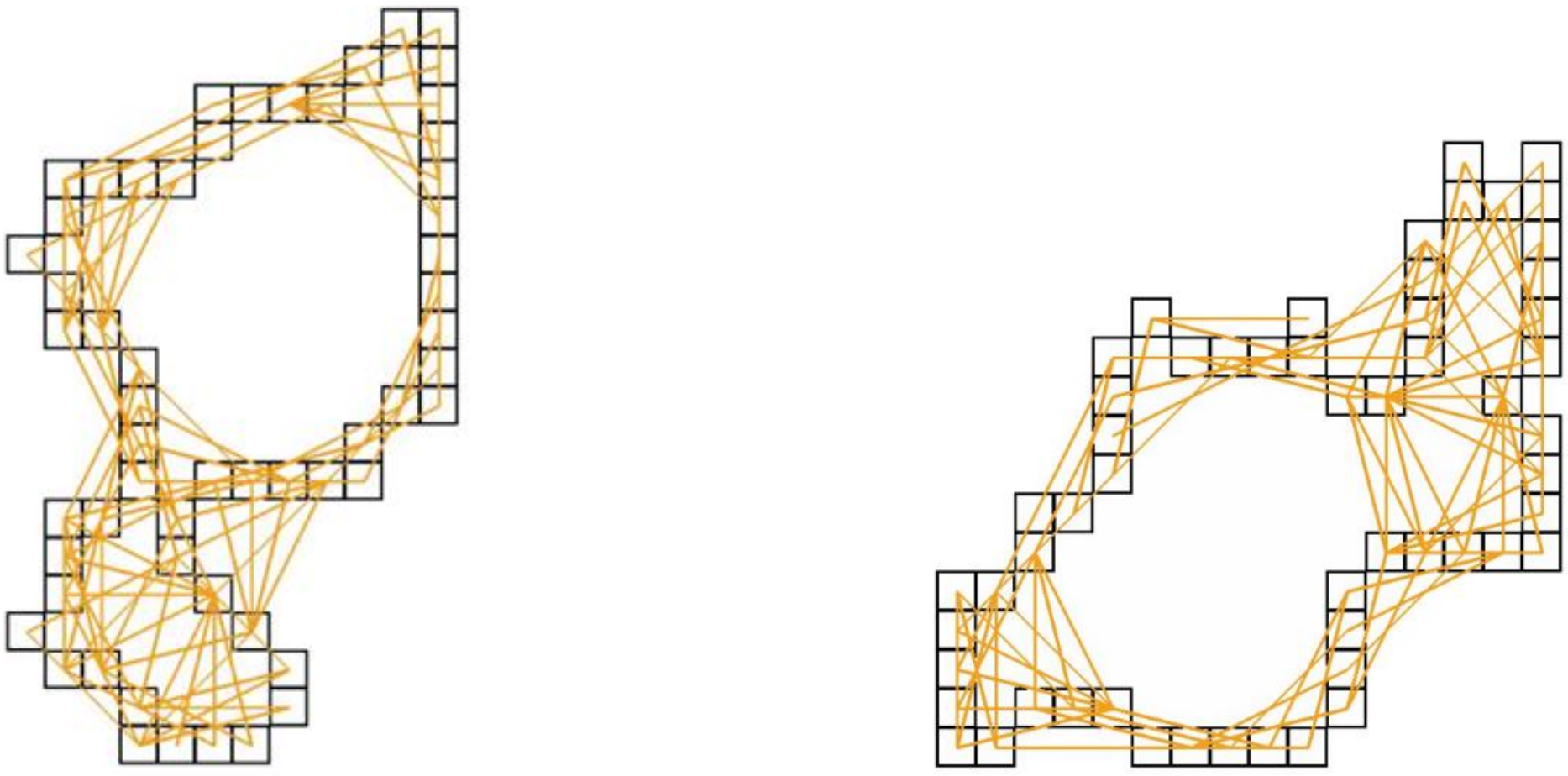

(**a**) (**b**)

**Figure 2.** The two collections of distances, marked by the orange lines, within $[i = 4, i+1 = 5)$ between the edge pixels are presented. (**a**) Edge of the chosen target inclusion with $h_{\text{target}}(4) = 95$ connections; (**b**) Edge of the related synthetic cluster with $h(4) = 87$ connections. Only the edge pixels affecting the value of the related interfaces are sketched. Both clusters have area = 143 and interface = 76, so their shape index $q \cong 0.157$ (the definition of $q$ used is given in the text).

For clarity, Figure 3 shows relevant histograms of the linear distances between the edge pixels for the above clusters.

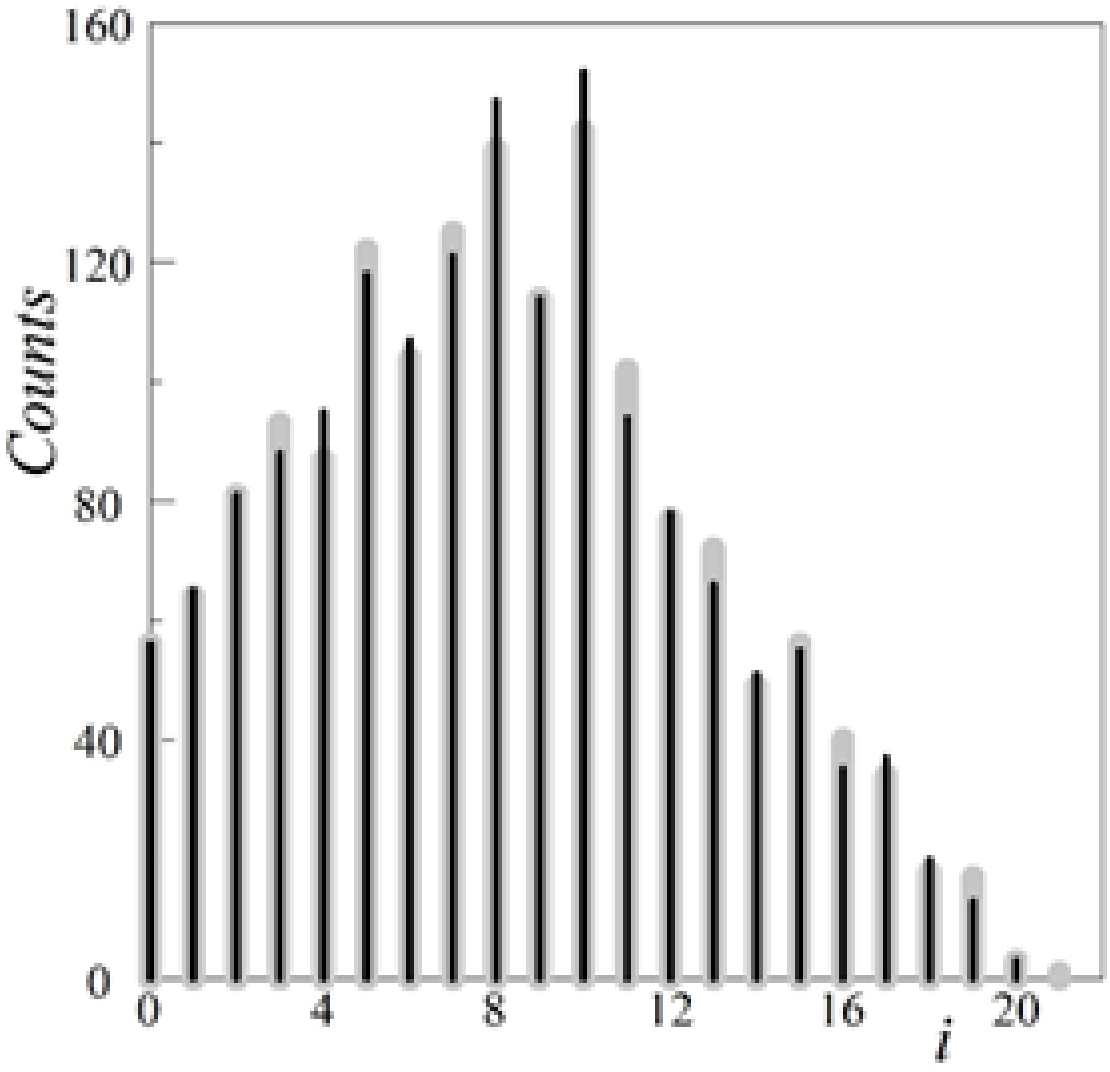

**Figure 3.** The histograms of linear distances between the edge pixels, $h_{\text{target}}(i)$, the drop black lines, and $h(i)$, the drop grey lines, for the chosen target inclusion Figure 2a, and the related synthetic cluster from Figure 2b, respectively. The common value $h_{\text{target}}(i = 0) = h(i = 0)$ corresponds to the number of the edge pixels.

It should be emphasized that the procedure for generating a basic set of synthetic clusters corresponding to a given set of target inclusions provides the same dimensionless shape index defined as $q$ = (cluster area)$^{1/2}$/(interface of cluster) in each pair containing real inclusion and its surrogate cluster. For instance, given any square or circle, the shape index is exactly 0.25 and about 0.282, respectively. In this way, the procedure ensures that also the average shape index $<q>$ is equal for both sets. This type of compatibility did not occur in our earlier models for creating initial cluster configurations [52].

## 3. Reconstruction Using Entropic Descriptors–Stage Two

In this article, we present a statistical two-stage reconstruction of microstructure using multi-scale extended entropic descriptors (Appendix in Ref. [27]). These EDs are associated with statistical properties at different length scales of any spatial configuration of finite size objects. The method is based on the observation that the main entropic descriptor provides a quantitative multi-scale characteristics of the spatial non-uniformity for a binary pattern representing a two-phase microstructure. Intriguingly, one can obtain the so-called *phase* entropic descriptors, which determine the spatial inhomogeneity attributed to each phase-component, making use of the decomposable multi-phase entropic descriptor [54]. More details with reference to binary patterns are given in the Appendix A while those for grey-level patterns can be found in [24,55].

The hybrid pair, $\{S_\Delta(k), C_S(k)\}$, is often used for multi-scale statistical reconstruction via entropic descriptors [24,51,52]. The first descriptor, $S_\Delta(k) \equiv [S_{max}(k) - S(k)]/\lambda(k)$, is a quantitative measure of the degree of spatial inhomogeneity per cell [21,24,48,49]. Here, $S(k)$ is the current entropy for the real configuration, and $S_{max}(k)$ denotes the entropy for the theoretically most homogeneous system, while $\lambda(k) = [L - k + 1]^2$ describes the number of partially overlapping sampling cells $k \times k$ for a given length scale $k$. In turn, the second descriptor $C_S(k) \equiv S_\Delta(k)\gamma(k)$ can be applied as a measure of the spatial statistical complexity [50], where the shortcut is used, $0 < \gamma(k) \equiv [S(k) - S_{min}(k)] / [S_{max}(k) - S_{min}(k)] < 1$. A better structural accuracy of statistical reconstructions can be expected using a set of three functions, $S_\Delta(k)\gamma^\alpha(k)$ with $\alpha = 0$, 1 and 2, which we will call extended ED-triplet, $\{S_\Delta(k), S_\Delta(k)\gamma(k), S_\Delta(k)\gamma^2(k)\}$ [27].

Some changes must be made to the preferred SA technique to make it compatible with our two-stage method for clusters, not pixels. First, the cluster SA technique should consider the modified cost function associated with the extended ED triplet. The modified objective function $E$ is treated here as the average "energy" per the descriptor curve and the length scale. The energy $E$ is related to the quantity commonly used in statistics as a formal goodness of fit test [45]. Here, the cost multi-scale function is the sum of the squared and normalized differences between the respective ED functions for current configuration and target pattern. The differences are normalized with respect to maximum target values. The corresponding simple formulas can be saved as

$$\tilde{S}_\Delta \gamma^\alpha - \tilde{S}_{\Delta,\mathrm{target}} \gamma^\alpha_{\mathrm{target}} \equiv [S_\Delta \gamma^\alpha - S_{\Delta,\mathrm{target}} \gamma^\alpha_{\mathrm{target}}] \big/ \max S_{\Delta,\mathrm{target}} \gamma^\alpha_{\mathrm{target}} , \tag{3}$$

and

$$E = \frac{1}{3n} \sum_{k}^{L} \sum_{\alpha=0}^{2} \left[ \tilde{S}_\Delta(k)\gamma^\alpha(k) - \tilde{S}_{\Delta,\mathrm{target}}(k)\gamma^\alpha_{\mathrm{target}}(k) \right]^2 . \tag{4}$$

For a given tolerance value, this combined objective function improves the structural accuracy of the reconstruction compared to the case where only one pair with one target curve was used. In particular, in Section 4, we consider a target sample of linear size $L = 170$ (in pixels). To reduce computational costs, we use every second scale of length $k = 2, 4, \ldots, 170$ in Equation (4). Thus, the actual number of analyzed length scales equals $n = 85$. After creating a set of $\{M\}$ initial random configurations consisting of synthetic clusters (previously prepared according to the procedure described in Section 2), the value $E(M)$ of the cost function in Equation (4) is calculated for each of these configurations. Then the configuration $m_i \in \{M\}$ with the lowest value $E(m_i)$ is selected as the initial one. This ensures a good starting position for the next step.

As mentioned earlier, we adapted the SA technique in these studies to use it for entire synthetic clusters. Instead of swapping pixels in different colors, each attempt concerns a synthetic cluster picked at random. It can move in a random direction and at a distance within the allowable range of step length if the synthetic cluster is not in contact with any of the other clusters. This point is crucial to speed up the process of statistical reconstruction. Further steps are quite typical. For a given temperature, loop and current configuration with $E_{old}$ energy, the next configuration with $E_{new}$ energy (called the new state of the system) and generated by randomly changing the position of a selected synthetic cluster is accepted with probability $p(\Delta E)$ given by the Metropolis-MC acceptance rule [1]

$$p(\Delta E) = \min[1, \exp(-\Delta E/T)], \tag{5}$$

where $\Delta E = E_{new} - E_{old}$ is the change in the energy between two successive states. After acceptance, the trial configuration becomes a current one, and the procedure is repeated. A quite aggressive cooling schedule, $T(l)/T(0) = (0.82)^l$, was used for temperature loops of increasing length [52]. Here $l$ numerates the loops, the initial fictitious temperature is $T(0) = 5 \times 10^{-5}$, and for the chosen tolerance value $\delta = 7 \times 10^{-5}$, we get sufficiently accurate reconstructions. In the next section, we will apply this method for a difficult case from the viewpoint of standard statistical reconstruction using the SA technique. Our method is particularly convenient to use for this type of aggregated isotropic materials that contain highly irregular inclusions.

## 4. Results

### *4.1. Statistical Reconstruction of Silica Dispersed in Rubber Matrix*

We would like to examine our approach on the Example_1 of a microstructure with 20.04% silica in a rubber matrix adapted from Bostanabad et al. (*cf.* Figure 8a of [34]). This type of two-phase material with branched inclusions of various sizes and irregular shapes poses a challenge for statistical reconstruction. In the cited article, the author introduces and tests a general methodology for characterization and reconstruction based on supervised learning that can be applied to a wide range of microstructures. For our purposes, we select a representative section 170 × 170 treated further as the target pattern, as shown in Figure 4a. It is clearly seen that the black phase is characterized by non-uniform spatial distribution of 113 irregular silica inclusions. Regarding the corresponding collections of synthetic clusters (we restricted ourselves to only three ones with different random generator seeds), they are easily obtained by the procedure described in Section 2; they can be found in the Supplementary Materials. Each of those collections characterizes the same average shape index per cluster as in the set of compact inclusions for a given target pattern $\langle q \rangle = \langle q_{target} \rangle = 0.199019$. Now, we create several random trial configurations, using the chosen collection of shapes, here with Seed 2. This allows selecting the one with the lowest initial energy value for the cost function defined by Equation (4). In the present approach, we construct the cost function using a family of three extended multi-scale entropic descriptors. Accordingly, Figure 4b shows the synthetic clusters configuration chosen as the initial one with the lowest value of the cost function, $E_{start} = 7 \times 10^{-3}$.

Now, we are ready to begin the second stage of the statistical reconstruction process. Using the SA program developed directly for clusters, we obtain an optimized microstructure corresponding to the first appearance of such a value of the cost function given by Equation (4), which ensures adequate reconstruction accuracy, as shown in Figure 4c. For our purposes, the appropriate tolerance value $\delta = 7 \times 10^{-5}$ is sufficient to obtain the energy value $E_{final} = 6.7 \times 10^{-5} < \delta$. In addition, two small frames (the orange online) focus our attention on a certain compact inclusion in Figure 4a and its corresponding synthetic cluster, as shown in Figure 4c.

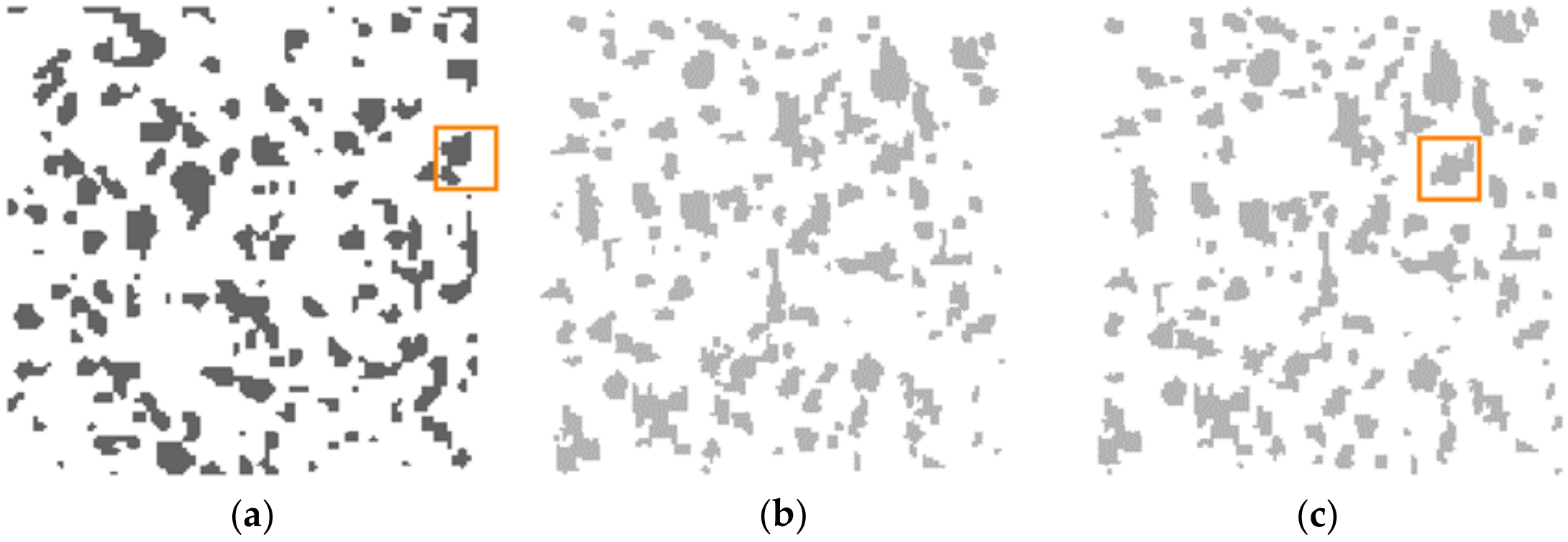

**Figure 4.** The images relate to the clustered microstructure. (**a**) Sub-domain 170 × 170 of the target pattern with silica ramified compact inclusions in a rubber matrix (adapted from [34]); (**b**) Initial configuration of synthetic clusters with the energy $E_{start} = 7 \times 10^{-3}$; (**c**) Optimized microstructure obtained under simulated annealing (SA) modified for clusters after 792 accepted MC-steps with $E_{final} = 6.7 \times 10^{-5} <$ tolerance $\delta = 7 \times 10^{-5}$.

In turn, Figure 5a,b shows the quality of the TSR. A comparison of the extended entropic descriptors $\{S_\Delta(k)\gamma^\alpha(k)\}$ with $\alpha = 0$, 1 and 2 for the target pattern (Figure 4a, the solid lines) with those for the optimized microstructure (Figure 4c, the open circles) is presented in Figure 5a. The results related to the initial configuration depicted in Figure 4b are represented by the dashed lines. The ED values were computed using hard-wall boundary conditions (HBC). The SA technique for clusters provides an optimized microstructure with sufficiently good fitting to the target curves after just 792 accepted MC-steps. This suggests the high efficiency of the proposed method of TSR for microstructures with arbitrarily shaped compact inclusions.

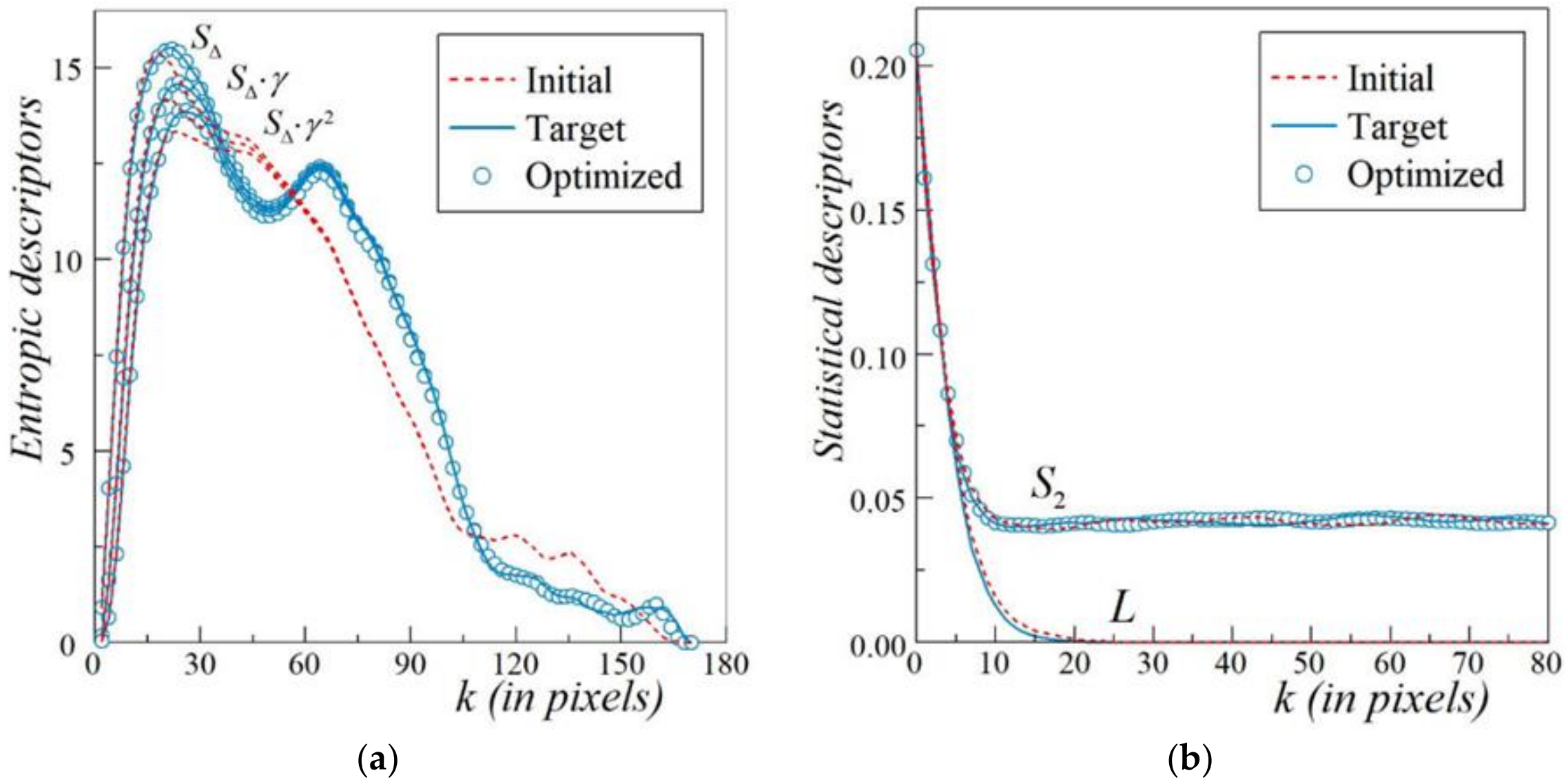

**Figure 5.** Accuracy of the two-stage reconstruction (TSR) of the microstructure with compact silica inclusions in a rubber matrix (adapted from [34]) supported by a comparison of the values of related initial, target and optimized functions. (**a**) Extended entropic descriptor (ED) triplet that is $S_\Delta(k)\gamma^\alpha(k)$ with $\alpha = 0$, 1 and 2, hard-wall boundary conditions (HBC) are imposed in the $x$ and $y$ directions; (**b**) $S_2(k)$, periodic boundary conditions (PBC) are imposed in both directions and the orthogonal lineal-path function $L(k)$ for HBC.

Additionally, Figure 5b reveals the interesting confirmation of the quality of our TSR. When two-point correlation function $S_2(k)$ and orthogonal lineal-path function $L(k)$ are computed for the initial configuration, the microstructure of target and the finally optimized pattern by mean of TSR, the corresponding values show a satisfactory agreement. Note that within HBC, the values of $L(k)$ calculated for the initial and optimized configuration are the same. Thus, only the dashed and solid lines are compared for this function.

*4.2. Statistical Reconstruction of Stones Dispersed in a Cement Paste*

For comparison, we test the capability of our approach on the Example_2 of realistic material. This concrete microstructure was originally reconstructed via a hybrid pair of two-point correlation function $S_2$ and cluster function $C_2$ in Jiao et al. (*cf.* Figure 2 of [47]). Such a two-phase material with relatively large and densely dispersed inclusions of various sizes and irregular shapes poses a challenge to statistical reconstruction. Figure 6a shows 170 × 170 cross-section for a concrete sample adapted from Ref. [47] as our target pattern. The 51.24% black phase represents 42 irregular stones randomly placed in the white phase, i.e., cement paste. The corresponding three collections of synthetic clusters generates the procedure described in Section 2 with the average shape index per cluster $<q> = <q_{\text{target}}> = 0.197867$; they can be found in the Supplementary Materials. From several random trial configurations of the synthetic clusters taken from the collection with Seed 3, we choose the one with the lowest initial energy value for the cost function defined by Equation (4). Accordingly, the synthetic clusters configuration chosen as the initial one with the lowest value of the cost function, $E_{\text{start}} = 6.2 \times 10^{-2}$ is presented in Figure 6b.

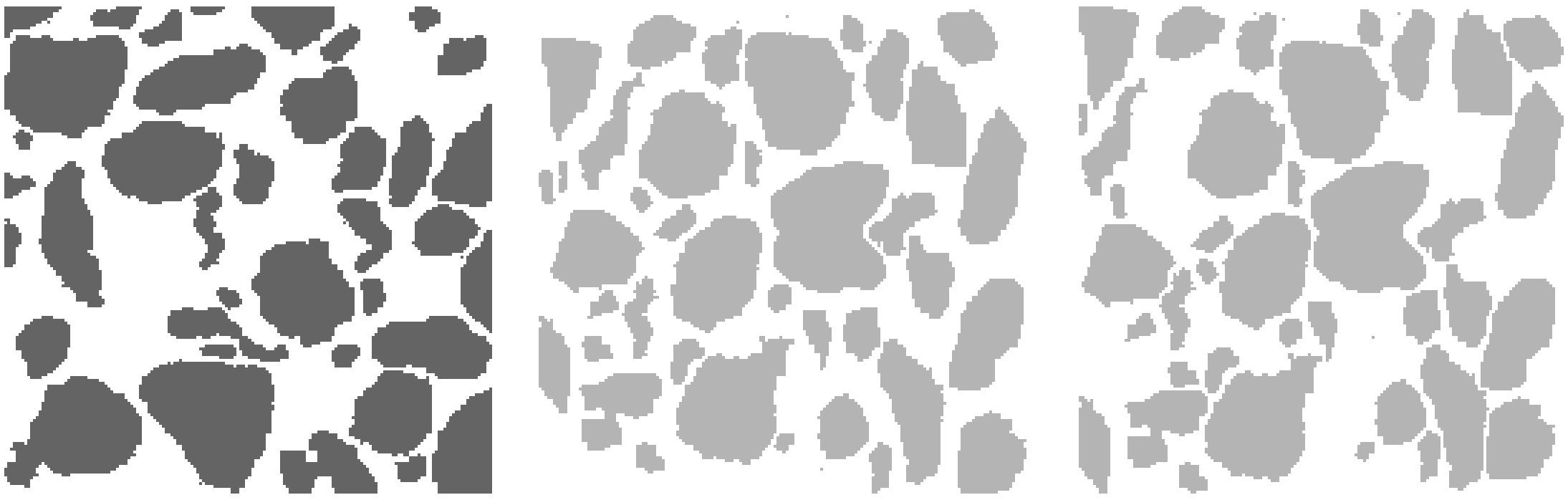

(**a**) (**b**) (**c**)

**Figure 6.** The images refer to the microstructure of the cross-section of the concrete sample. (**a**) Sub-domain 170 × 170 of the target pattern with stones dispersed in cement paste (adapted from [47]); (**b**) Initial configuration of synthetic clusters with the energy $E_{\text{start}} = 6.2 \times 10^{-2}$; (**c**) Optimized microstructure obtained under SA modified for clusters after 483 accepted MC-steps with $E_{\text{final}} = 6.8 \times 10^{-5} <$ tolerance $\delta = 7 \times 10^{-5}$.

The quality of TSR is confirmed by Figure 7a,b. A comparison of the extended entropic descriptors $\{S_\Delta(k)\gamma^\alpha(k)\}$ with $\alpha = 0$, 1 and 2 for the target pattern (Figure 6a, the solid lines) with those for the optimized microstructure (Figure 6c, the open circles) is presented in Figure 7a.

On the other hand, this binarized concrete sample cross-section [47] was stochastically reconstructed by Tahmasebi and Sahimi [56], and Olchawa and Piasecki [52]. Respective results of comparable quality were obtained using completely different methods. In the former paper, a cross-correlation function was applied (*cf.* Figure 4 of [56]). Focusing on the latter case, the so-called weighted double-hybrid (WDH) method was utilized with parameters: slightly more demanding tolerance $\delta = 10^{-5}$, similar cooling schedule $T(l)/T(0) = (0.85)^l$ and a lower initial temperature $T(0) = 10^{-7}$ (*cf.* Figure 5 of [52]). That approach uses simultaneously a hybrid pair of entropic descriptors $\{S_\Delta(k), C_S(k)\}$ having a weighting factor $\alpha$, (this symbol in the present paper has a different meaning), and a pair of correlation functions $\{S_2, C_2\}$ with a weighting factor $(1-\alpha)$. It is worth adding one correcting remark. Formula (3.3) of [52], which describes the final form of the cost function, lacks the weighting factor $\alpha$ due to a printing error, but this does not affect the conclusions.

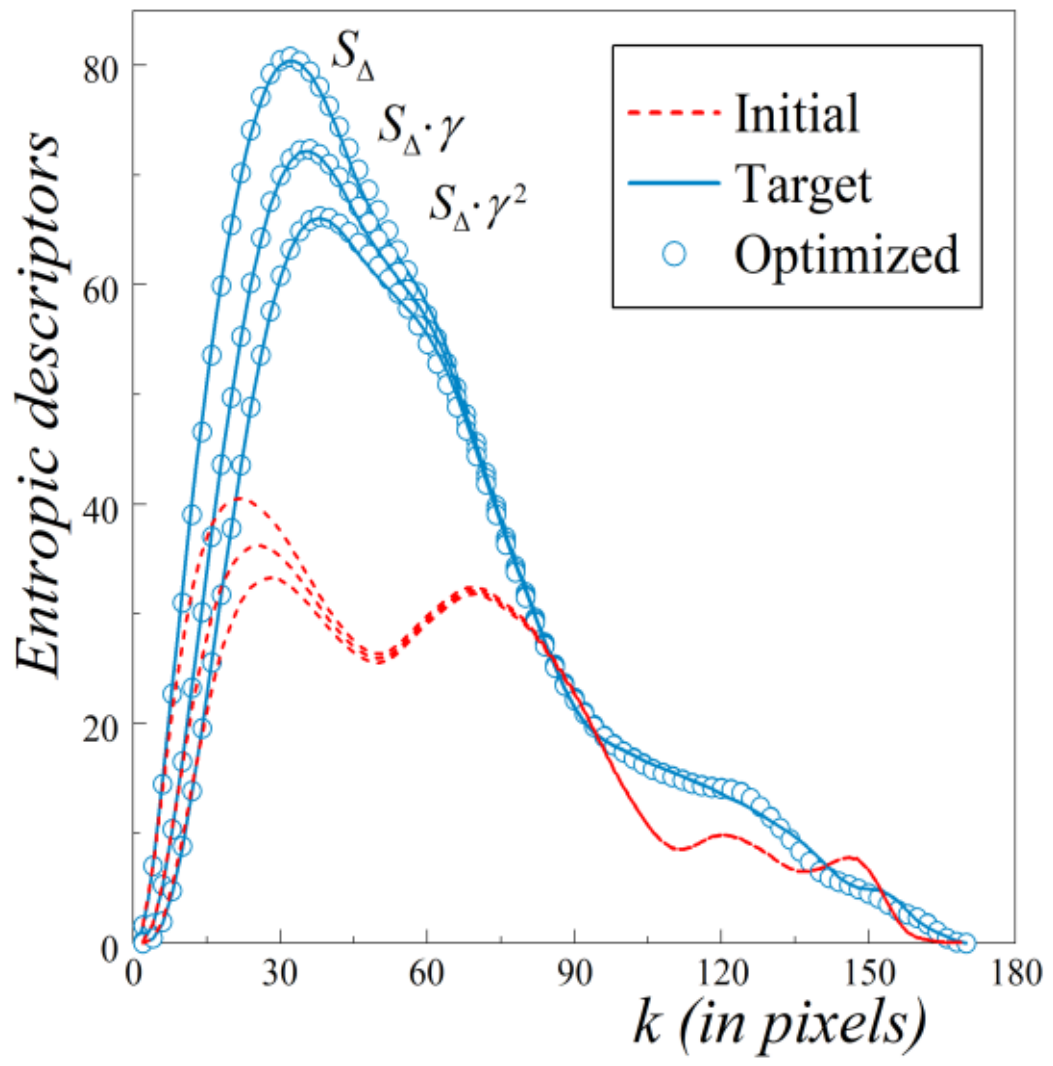

(**a**)

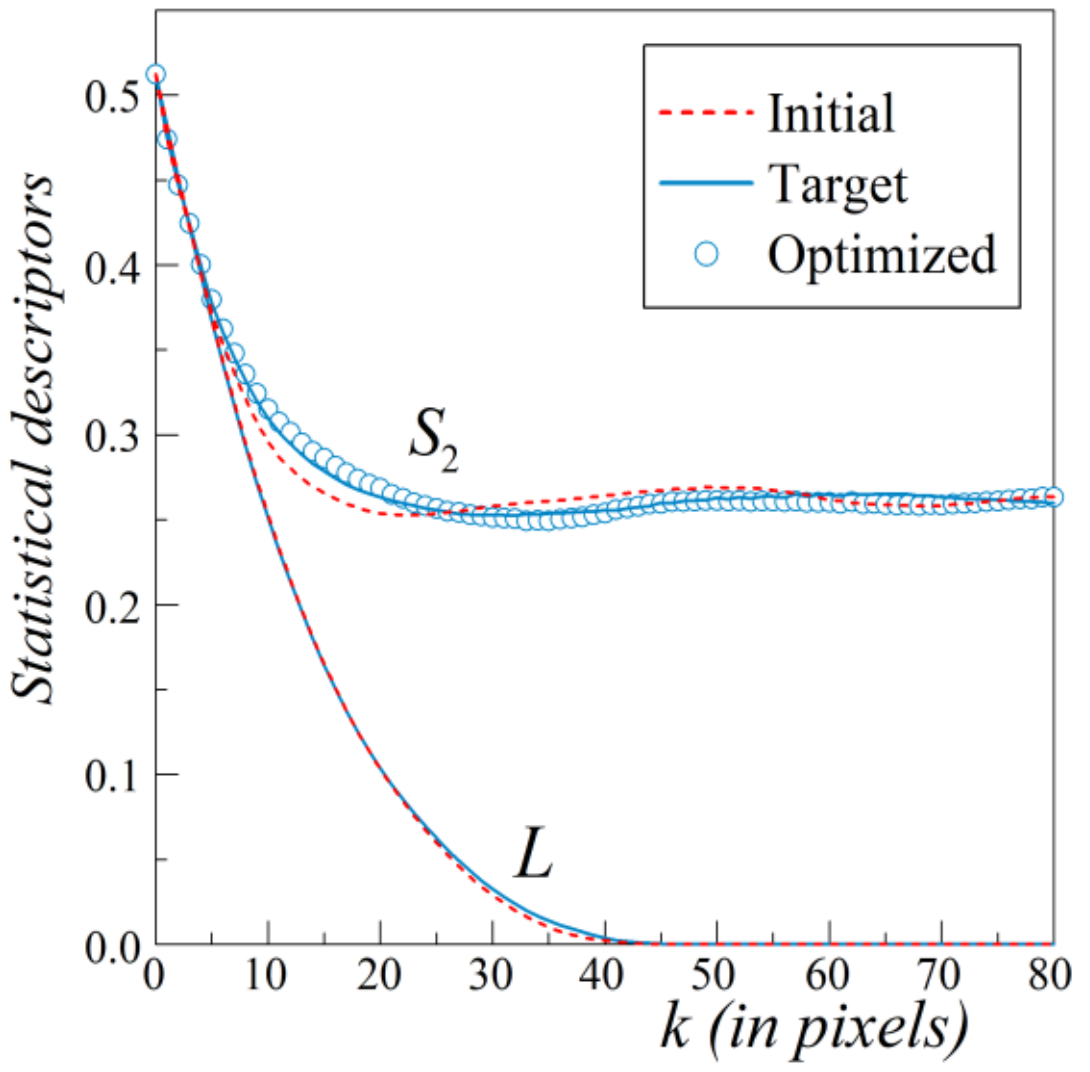

(**b**)

**Figure 7.** Accuracy of the TSR of the microstructure with stones in a cement paste (adapted from [47]) supported by a comparison of the values of related initial, target, and optimized functions. (**a**) Extended ED triplet that is $S_\Delta(k)\gamma^\alpha(k)$ with $\alpha = 0$, 1 and 2, HBC are imposed in the *x* and *y* directions; (**b**) $S_2(k)$, PBC are imposed in both directions and the orthogonal lineal-path function $L(k)$ for HBC.

This is clearly seen in (Figure 8 of [52]), which shows the convergence of energy *E* as a function of the accepted Monte Carlo attempts $N_a$ for three representative cases $\alpha = 0.9$, 0.5, and 0.1 that the corresponding final $N_a(\alpha = 0.1)$ value is the highest (about 5800). On the other hand, taking into account the present tolerance value $\delta = 7 \times 10^{-5}$, one can estimate the $N_a(\alpha = 0.1) \cong 2000 > 483$. It means a much larger number of MC steps than the present one accepted in the TSR method. The larger the correlation functions component (i.e., the smaller the $\alpha$ is), the higher the number of the finally accepted MC steps that is needed (*cf.* Figures 3 and 7 of [52]). The following observation can therefore be given: the computational cost of the TSR method (measured with the required MC steps) compared to the reconstruction method using two-point spatial correlation functions is much lower.

This allows us to expect a high performance of the proposed TSR method for microstructures with even large compact inclusions. Similarly as in the previous example, the additional confirmation of the quality of the TSR approach can be found in Figure 7b. The corresponding $S_2(k)$ and $L(k)$ values computed for the relevant patterns show a satisfactory agreement.

## 5. Discussion

Since the values of $L(k)$ do not depend on the spatial distribution of these clusters, it seems reasonable to carry out the preliminary selection among a few generated sets of synthetic clusters. For example, as the preferred basic set, we can choose the one (*) for which we obtain the smallest value of the sum over *k* for $[L^*(k) - L_{target}(k)]^2$. Only now do we use Equation (4) to select the minimum value of the cost function $E(M^*)$ calculated for each of the prepared $M^*$ random configurations of synthetic clusters belonging to the preferred set (*). This initial configuration would be optimal at the second stage for future applications of modified SA for clusters.

It is noteworthy that the SA modified for clusters shows a significant advantage over other methods–which use a random initial configuration of pixels–in particular, when the target microstructure contains clusters relatively large in comparison with the sample size. This observation is suggested by Example_2 where the results obtained for a sample of concrete microstructure reconstructed were already analyzed in [47,52,56].

On the other hand, the current TSR approach can also be adapted to the reconstruction of porous microstructures provided we know at least the complete interface. It should be emphasized that the areas and interfaces of the target inclusions are accurately reproduced by synthetic clusters, but their shapes differ in the stochastic context. The final planar configuration of synthetic clusters is optimized within the modified SA. The SA algorithm for clusters can also be extended to allow grain-to-grain contact, e.g., for low-density systems even with strong polydispersity in grain sizes. However, with this approach, some fluctuations in interface values are inevitable.

It should be remembered that the information contained in the specified set of entropic descriptors might also represent a hypothetical medium based on simple models. Then, the problem of generating of the underlying microstructure is frequently referred to as a "construction" [45]. In this way, it is also possible to produce a kind of model granular structures with highly irregular shapes of granules dispersed in a dense, sticky medium by the TSR method. Finally, the approach can also be adapted to the more general case of non-compact clusters. However, research is currently underway on the three-dimensional version of the TSR method.

Shortly after this study had been completed, the authors became aware of the existence of a conceptually similar article of You et al. [57]. The algorithm proposed there is a three-step process that generates statistically equivalent representative volume elements (RVE) of the size $150 \times 150 \times 150$ (in pixels) from the particle shape repository. This repository reflects realistic microstructures obtained from micro-computed tomography scanning of mechanoluminescent material. At the final stage, the difference between the histograms of the targeted and reconstructed cumulative distribution function is minimized by the SA technique. Using the $S_2(r)$ for verification of the method, You et al. show Figure 10 in [57] with two diagrams (N.B. the corresponding descriptions, $\Phi_{11}$ and $\Phi_{12}$, of ordinate axes should be interchanged). According to them, a slight gap between the reconstructed and the target model occurring in the characteristic range of distances, $10 < r < 30$, could be explained with the stochastic characteristics of the reconstruction and small differences of particle sizes.

Here, despite the not very high tolerance, $\delta = 7\times10^{-5}$, assumed in the TSR method, this effect for $S_2(k)$ is hardly visible in practically the entire range of length scales, cf. Figure 5b for Example_1 and Figure 7b for Example_2. This is a pleasant and side feature because the cost function in the TSR method does not use any correlation functions.

## 6. Conclusions

We offer an innovative two-stage method that can be used to generate statistically equivalent synthetic microstructures. This approach uses selected length scale-dependent entropic descriptors. Particular emphasis was placed on binary microstructures with arbitrarily shaped inclusions. For a given binary target pattern, the proposed method for reconstructing any heterogeneous microstructure with truly irregular inclusions is divided into two basic stages. At the first stage, a set of synthetic clusters is created, whose number corresponds to the target pattern. The procedure used ensures the same clusters areas and interface values as in the target pattern. This way, individual shape indices as well as their average value per cluster are also preserved. Thus, some significant statistical characteristics of the target pattern are taken into account. At the second stage, the approach uses the simulated annealing technique adapted to clusters. As a result, the computational cost in the two-stage method is much lower compared to the standard approach. Our method has been tested on adapted samples of irregular silica inclusions in a rubber matrix as well as stones of various sizes dispersed in a cement paste. Final statistical reconstructions suggest that even for morphologically complex material, this low-cost method easily generates statistically reliable synthetic patterns. Due to the packaging problem, with unusual or branched shapes of inclusions, the second stage of our method is effective when the concentration of inclusions is less than 0.5, see the Example_1, or slightly exceeds this value as in the Example_2. In addition, two-point correlation functions calculated for comparison purposes for a given reconstruction and target pattern are in satisfactory agreement. This further confirms the flexibility of our approach. We believe that the method presented here can be useful for the fast and low-cost reconstruction of a wide range of cluster microstructures with arbitrarily shaped grains.

**Author Contributions:** Conceptualization, R.P., W.O., and D.F.; Methodology, R.P., W.O., D.F., and A.B.; Software, W.O. and D.F.; Investigation, R.P., W.O., and D.F.; Graphics, D.F. and A.B.; Data collection, A.B.; Writing, R.P. with contribution W.O. and D.F. All the authors have read and agreed to the published version of the manuscript.

**Conflicts of Interest:** The authors declare no conflict of interest.

## Appendix A

For better readability, we briefly present the most important details related to the binary entropy descriptors defined for black pixels, which are treated here as unit objects. The EDs make use of micro-canonical configurational entropy $S(k) = \ln \Omega(k)$, its maximum possible value, $S_{max}(k) = \ln \Omega_{max}(k)$, as well as the minimum one, $S_{min}(k) = \ln \Omega_{min}(k)$. Here, the Boltzmann constant $k_B = 1$. The length scale is determined by the side length of a square sampling cell of the size $k \times k$ sliding by a unit lattice constant. Thus, the number of allowed positions for the sliding cell is equal to $\lambda(k) = [L - k + 1]^2$, which leads to a set of cell occupation numbers $\{n_i(k)\}$, $i = 1, 2, \ldots, \lambda(k)$. To simplify the notation, we skip the symbol $k$ whenever possible. "Maps" composed of sampled cells placed in a non-overlapping manner can be regarded as representative because they clearly reproduce on each scale $k$ the overall structure of the real initial sample. Such an approach allows computing the actual entropy $S(k)$ related to the configurational actual macrostate AM($k$) $\equiv \{n_i \in (0, 1, \ldots, k^2)\}$ realized by the number of microstates

$$\Omega(k) = \prod_{i=1}^{\lambda} \binom{k^2}{n_i}. \tag{A1}$$

The reference maximum possible value $S_{max}(k)$ is associated with the most uniform macrostate $\mathrm{RM}_{max}(k) \equiv \{n_i \in (n_0, n_0 + 1)\}_{max}$, with $\lambda - r_0$ number of cells occupied by $n_0 \in (0, 1, \ldots, k^2 - 1)$ and $r_0$ cells populated by $n_0 + 1$ of black pixels. The number of $\mathrm{RM}_{max}$-realizations is given by

$$\Omega_{max}(k) = \binom{k^2}{n_0}^{\lambda - r_0} \binom{k^2}{n_0 + 1}^{r_0}. \tag{A2}$$

Here, $r_0 = N \bmod \lambda$ and $n_0 = (N - r_0)/\lambda$. In turn, the reference minimum possible value $S_{min}(k)$ corresponds to the most spatially inhomogeneous macrostate $\mathrm{RM}_{min}(k) \equiv \{n_i \in (0, 0 < n < k^2, k^2)\}_{min}$, with $\lambda - q_0 - 1$ of empty cells, at the most one cell occupied with the number $n$ of pixels and $q_0$ of fully occupied cells. The obvious relation holds: $N = n + q_0 k^2$, where $n = N \bmod k^2$, $q_0 = (N - n)/k^2$ and $q_0 \in (0, 1, \ldots, \lambda - 1)$. The number of $\mathrm{RM}_{min}$-microstates can be written as

$$\Omega_{min}(k) = \binom{k^2}{0}^{\lambda - q_0 - 1} \binom{k^2}{n} \binom{k^2}{k^2}^{q_0} \equiv \binom{k^2}{n}. \tag{A3}$$

The entropic descriptor $S_\Delta \equiv [S_{max} - S]/\lambda$ quantifies the averaged per cell pattern's spatial inhomogeneity (a measure of configurational non-uniformity) by taking into account the average departure of a system's entropy $S$ from $S_{max}$ [48,49]. When a system's actual entropy $S \to S_{min}$, the spatial inhomogeneity becomes maximal. For a more detailed spatial analysis of a given binary pattern, the simplest hybrid approach using a pair $\{S_\Delta, C_S\}$ can be recommended. Here, the $C_S$ descriptor measures the so-called statistical spatial complexity [50]. The entropic descriptor $C_S \equiv S_\Delta \cdot \gamma$ is able to quantify the complex spatial behaviour by taking simultaneously into account the average departures of a system's entropy $S$ from both its maximum possible value $S_{max}$ and its minimum possible value $S_{min}$ [50]. This becomes clear after recalling the definition of the shortcut $\gamma = [S - S_{min}]/[S_{max} - S_{min}]$. If the two departures, $S_{max} - S$ and $S - S_{min}$, are comparable, then statistical complexity is maximal. Of course, we can use similar ideas to obtain grey-level equivalents of the above EDs, which can be useful for multi-phase materials [21,24,55].

## **Supplementary Material**: Shape libraries for Example_1

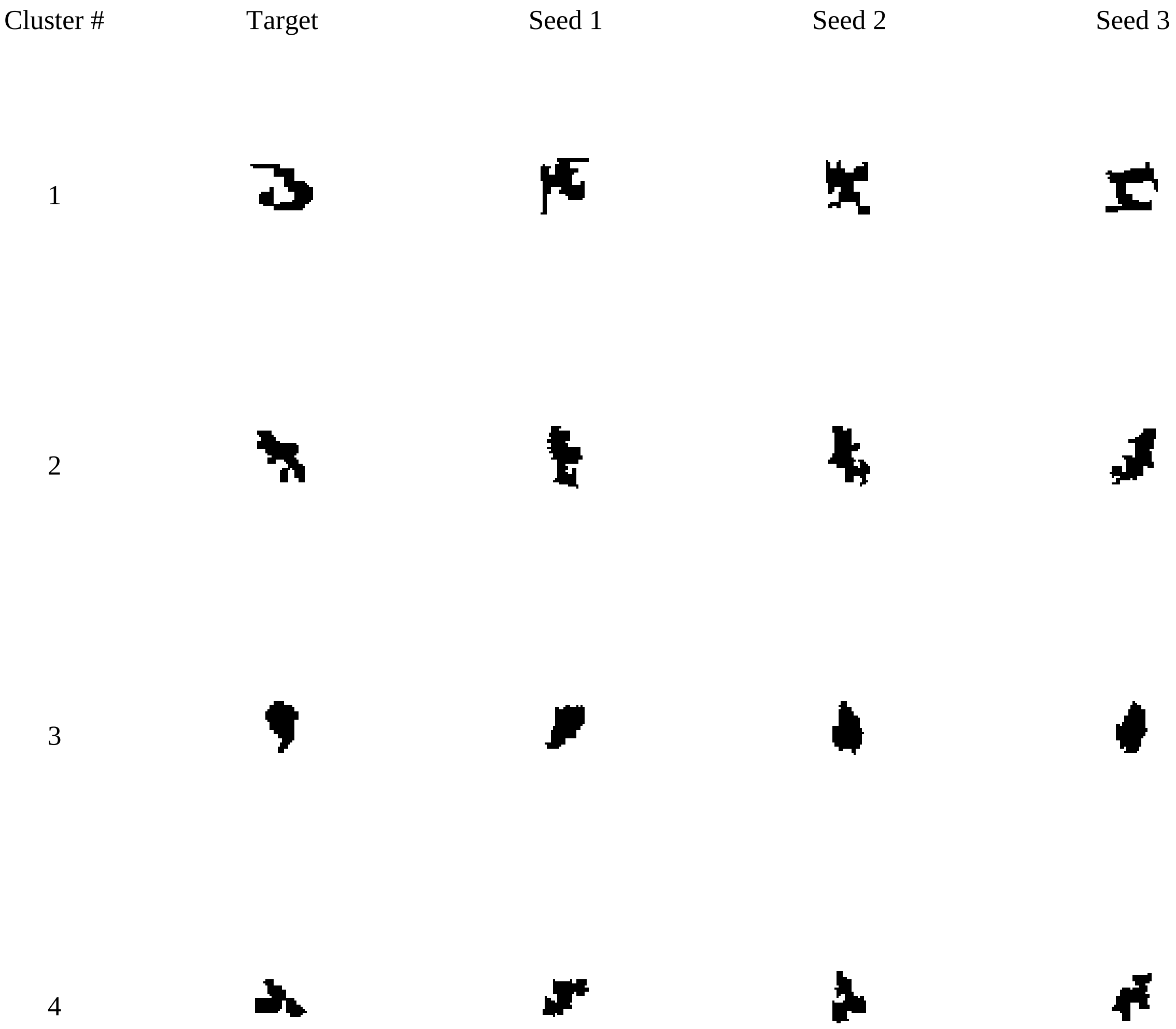

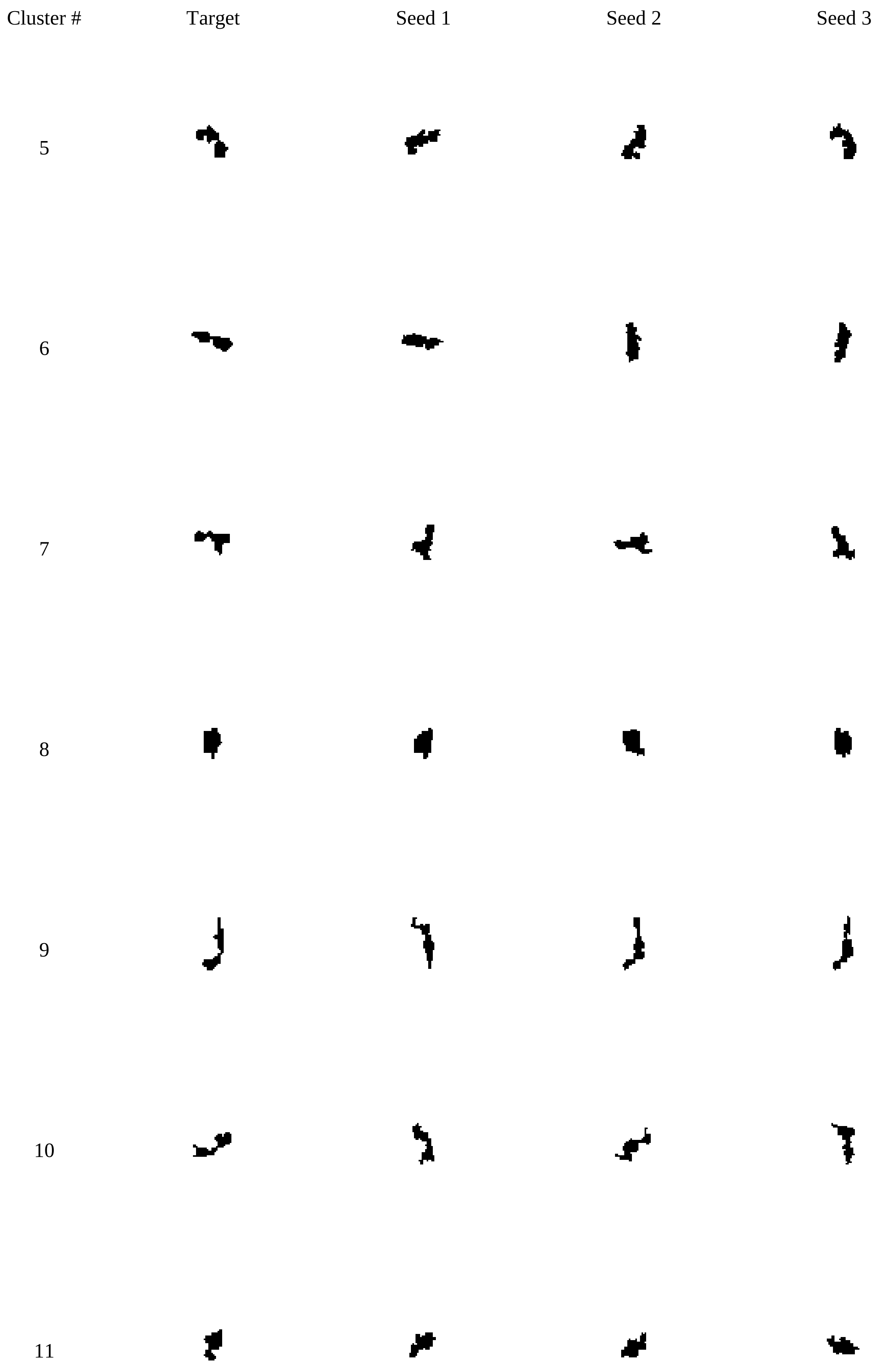

| Cluster # | Target | Seed 1 | Seed 2 | Seed 3 |
|---|---|---|---|---|
| 12 | 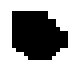 | 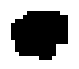 | 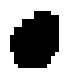 | 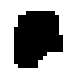 |
| 13 | 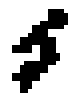 | 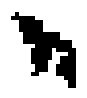 | 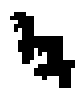 | 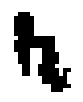 |
| 14 | 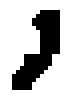 | 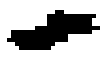 | 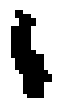 | 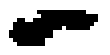 |
| 15 | 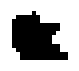 | 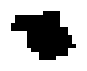 | 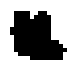 | 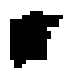 |
| 16 | 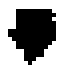 | 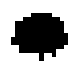 | 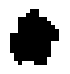 | 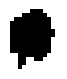 |
| 17 | 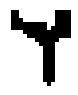 | 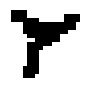 | 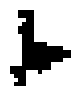 | 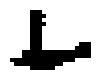 |
| 18 | 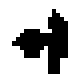 | 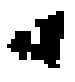 | 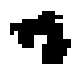 | 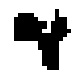 |

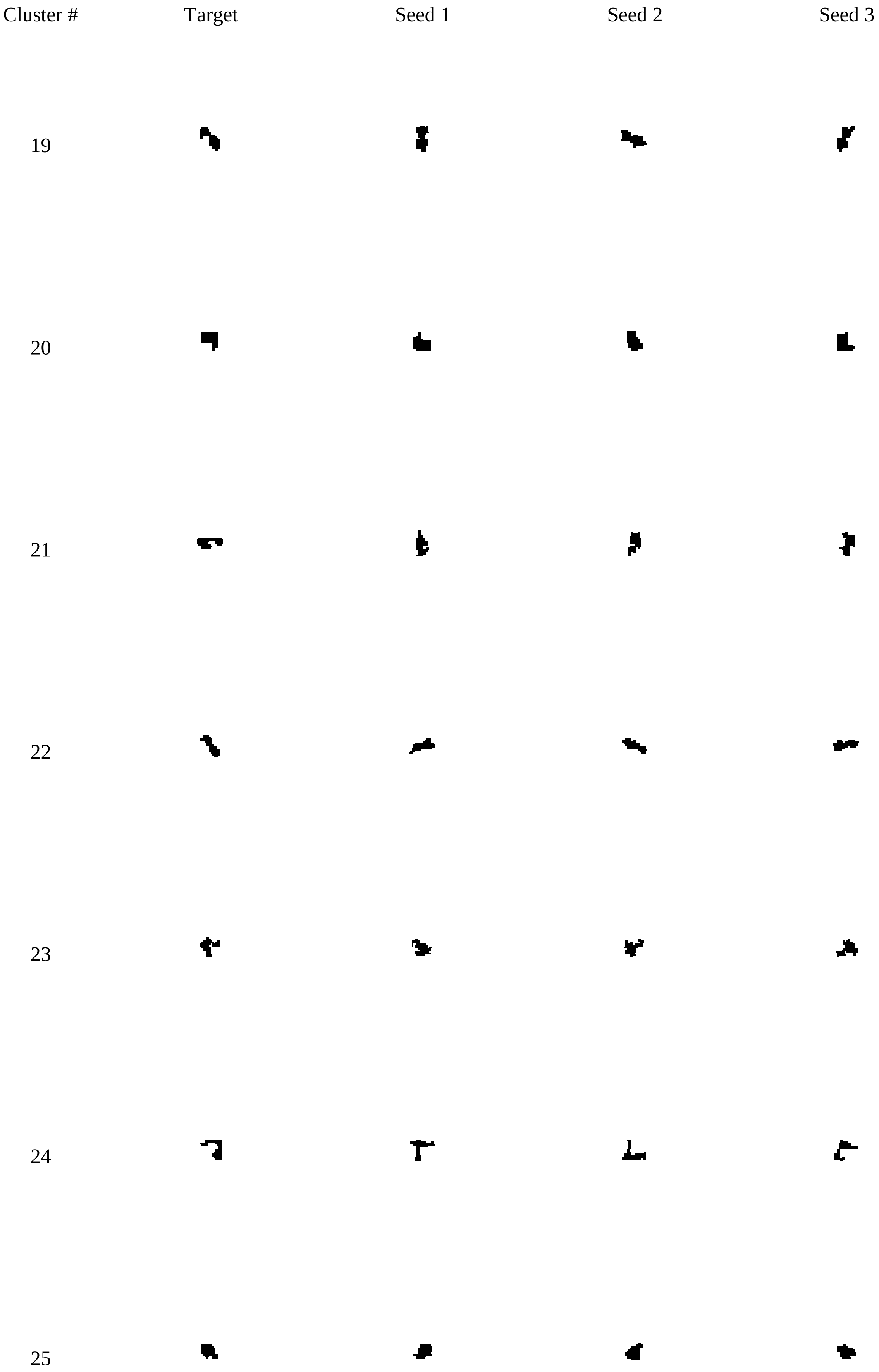

| Cluster # | Target | Seed 1 | Seed 2 | Seed 3 |
|---|---|---|---|---|
| 26 | 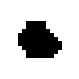 | 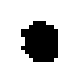 | 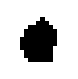 | 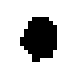 |
| 27 | 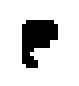 | 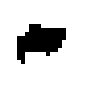 | 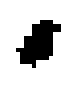 | 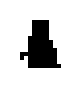 |
| 28 | 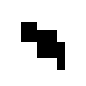 | 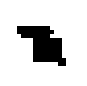 | 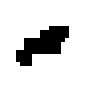 | 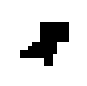 |
| 29 | 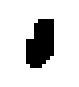 | 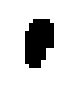 | 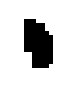 | 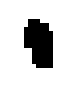 |
| 30 | 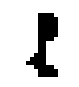 | 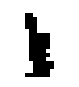 | 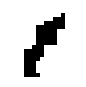 | 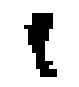 |
| 31 | 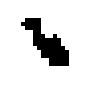 | 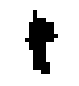 | 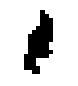 | 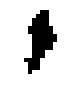 |
| 32 | 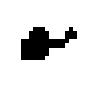 | 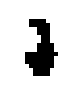 | 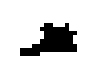 | 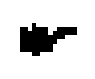 |

| Cluster # | Target | Seed 1 | Seed 2 | Seed 3 |
|---|---|---|---|---|
| 33 | 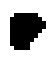 | 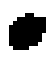 | 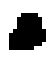 | 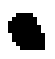 |
| 34 | 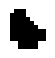 | 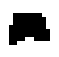 | 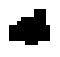 | 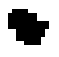 |
| 35 | 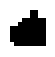 | 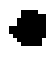 | 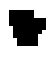 | 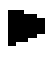 |
| 36 | 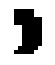 | 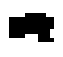 | 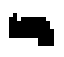 | 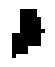 |
| 37 | 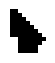 | 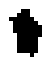 | 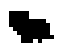 | 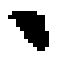 |
| 38 | 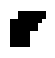 | 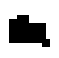 | 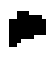 | 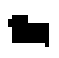 |
| 39 | 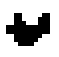 | 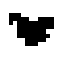 | 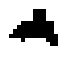 | 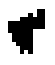 |

| Cluster # | Target | Seed 1 | Seed 2 | Seed 3 |
|---|---|---|---|---|
| 40 | 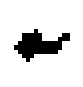 | 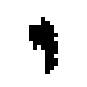 | 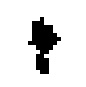 | 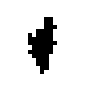 |
| 41 | 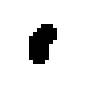 | 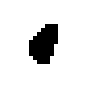 | 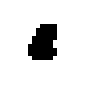 | 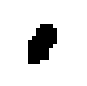 |
| 42 | 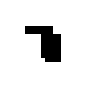 | 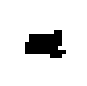 | 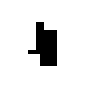 | 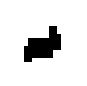 |
| 43 | 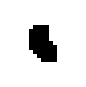 | 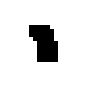 | 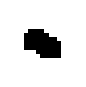 | 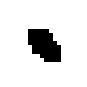 |
| 44 | 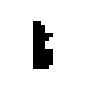 | 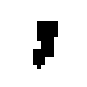 | 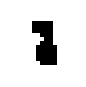 | 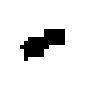 |
| 45 | 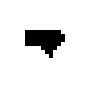 | 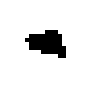 | 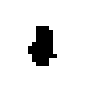 | 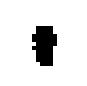 |
| 46 | 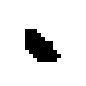 | 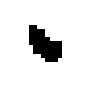 | 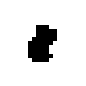 | 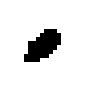 |

| Cluster # | Target | Seed 1 | Seed 2 | Seed 3 |
|---|---|---|---|---|
| 47 | 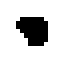 | 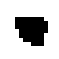 | 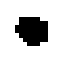 | 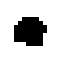 |
| 48 | 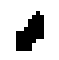 | 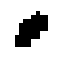 | 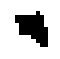 | 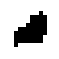 |
| 49 | 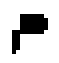 | 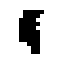 | 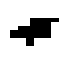 | 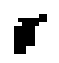 |
| 50 | 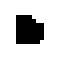 | 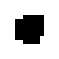 | 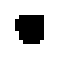 | 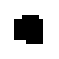 |
| 51 | 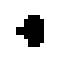 | 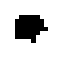 | 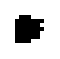 | 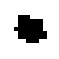 |
| 52 | 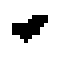 | 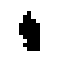 | 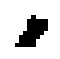 |  |
| 53 | 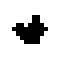 | 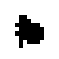 | 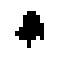 | 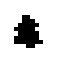 |

| Cluster # | Target | Seed 1 | Seed 2 | Seed 3 |
|---|---|---|---|---|
| 54 |  |  | 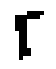 | 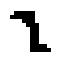 |
| 55 | 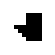 | 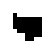 | 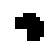 | 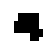 |
| 56 |  | 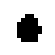 |  | 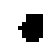 |
| 57 | 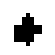 | 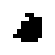 | 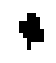 | 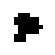 |
| 58 | 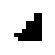 | 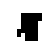 |  | 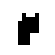 |
| 59 | 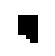 | 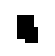 |  | 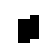 |
| 60 |  |  |  |  |

| Cluster # | Target | Seed 1 | Seed 2 | Seed 3 |
|---|---|---|---|---|
| 61 |  |  |  |  |
| 62 |  |  | 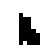 | 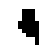 |
| 63 | 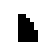 | 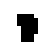 |  | 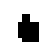 |
| 64 | 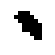 | 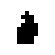 | 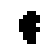 | 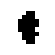 |
| 65 | 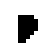 |  |  |  |
| 66 |  |  |  |  |
| 67 |  |  |  |  |

| Cluster # | Target | Seed 1 | Seed 2 | Seed 3 |
|---|---|---|---|---|
| 68 | 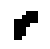 |  |  |  |
| 69 |  |  |  |  |
| 70 |  |  |  |  |
| 71 |  |  |  |  |
| 72 |  |  |  |  |
| 73 |  |  |  |  |
| 74 |  |  |  |  |

| Cluster # | Target | Seed 1 | Seed 2 | Seed 3 |
|---|---|---|---|---|
| 75 |  |  |  |  |
| 76 |  |  |  |  |
| 77 |  |  |  |  |
| 78 |  |  |  |  |
| 79 |  |  |  |  |
| 80 |  |  |  |  |
| 81 |  |  |  |  |

| Cluster # | Target | Seed 1 | Seed 2 | Seed 3 |
| --- | --- | --- | --- | --- |
| 82 | | | | |
| 83 | | | | |
| 84 | | | | |
| 85 | | | | |
| 86 | | | | |
| 87 | | | | |
| 88 | | | | |

| Cluster # | Target | Seed 1 | Seed 2 | Seed 3 |
|---|---|---|---|---|
| 89 | | | | |
| 90 | | | | |
| 91 | | | | |
| 92 | | | | |
| 93 | | | | |
| 94 | | | | |
| 95 | | | | |

| Cluster # | Target | Seed 1 | Seed 2 | Seed 3 |
|---|---|---|---|---|
| 96 | | | | |
| 97 | | | | |
| 98 | | | | |
| 99 | | | | |
| 100 | | | | |
| 101 | | | | |
| 102 | | | | |

| Cluster # | Target | Seed 1 | Seed 2 | Seed 3 |
|---|---|---|---|---|
| 103 | | | | |
| 104 | | | | |
| 105 | | | | |
| 106 | | | | |
| 107 | | | | |
| 108 | | | | |
| 109 | | | | |

| Cluster # | Target | Seed 1 | Seed 2 | Seed 3 |
|---|---|---|---|---|
| 110 |  |  |  |  |
| 111 |  |  |  |  |
| 112 |  |  |  |  |
| 113 |  |  |  |  |

**Supplementary Material**: Shape libraries for Example_2

| Cluster # | Target | Seed 1 | Seed 2 | Seed 3 |
|---|---|---|---|---|
| 1 | 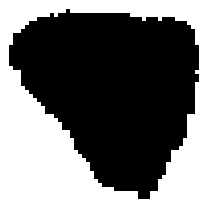 | 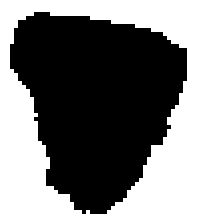 | 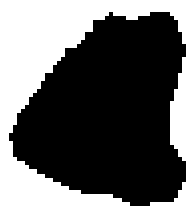 | 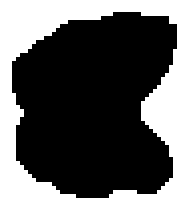 |
| 2 | 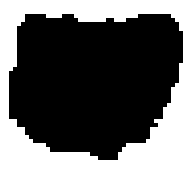 | 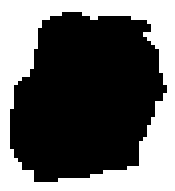 | 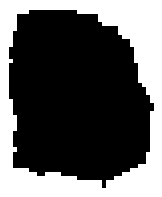 | 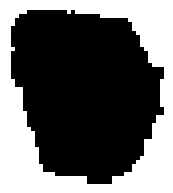 |

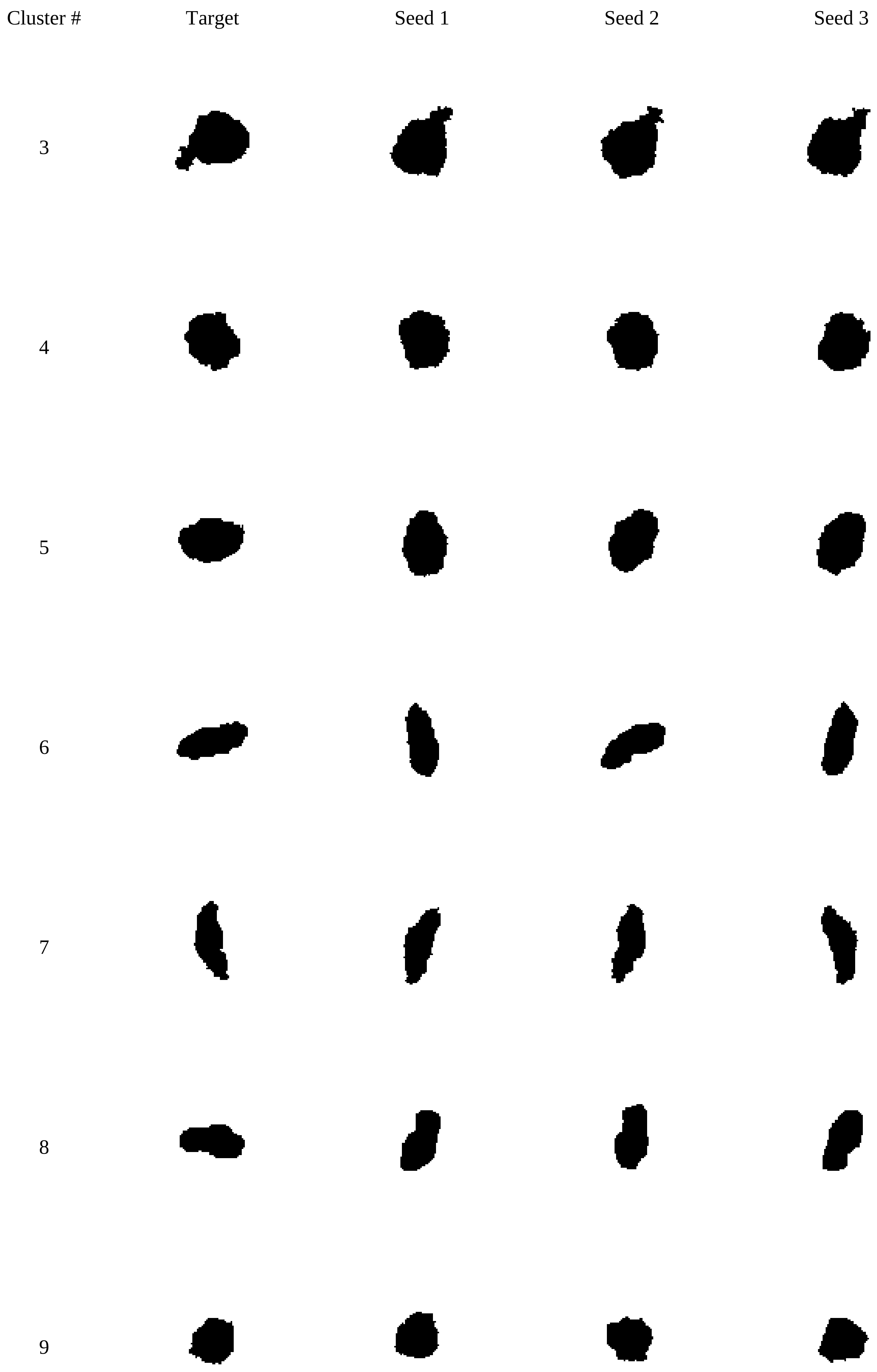

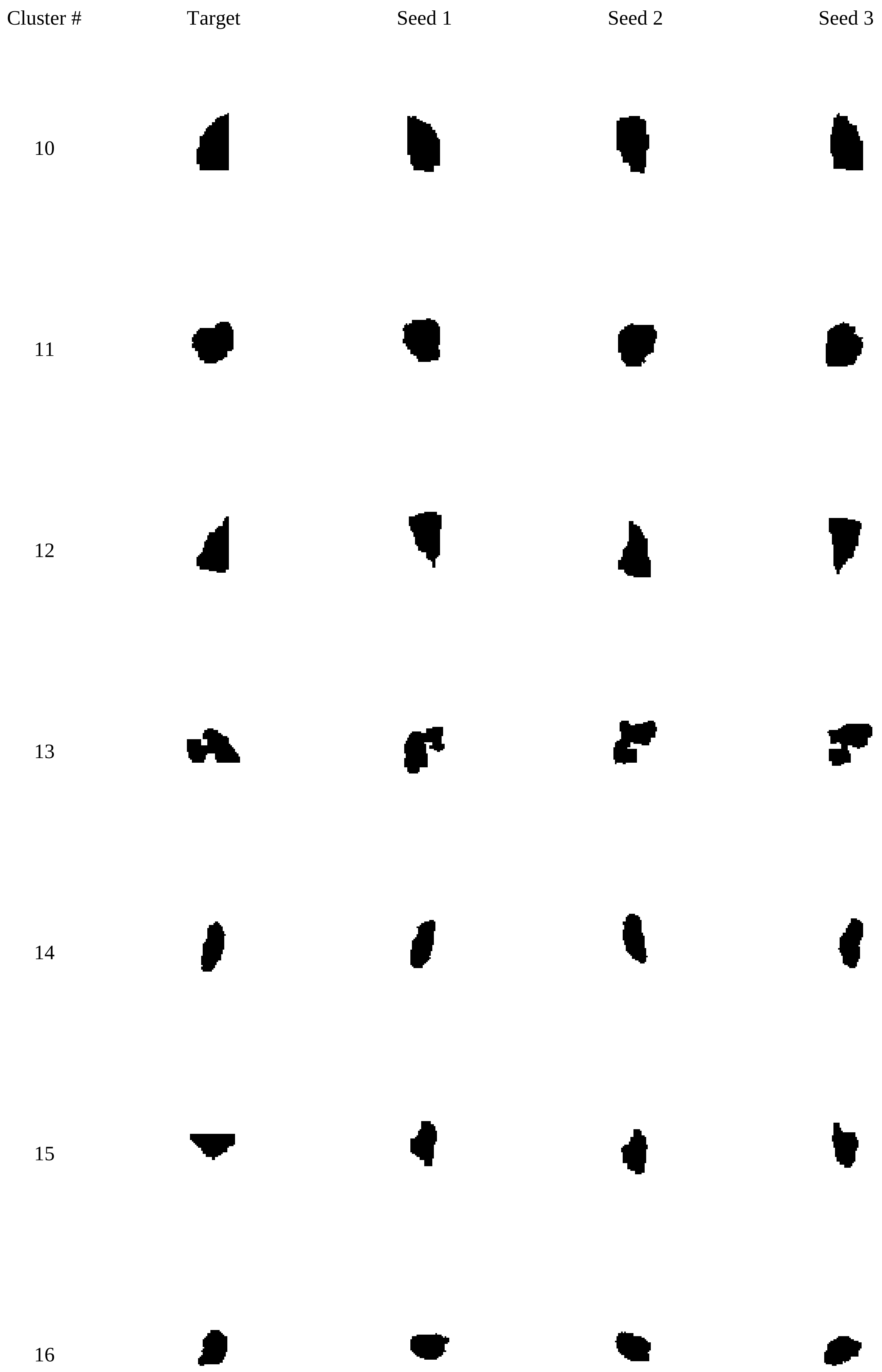

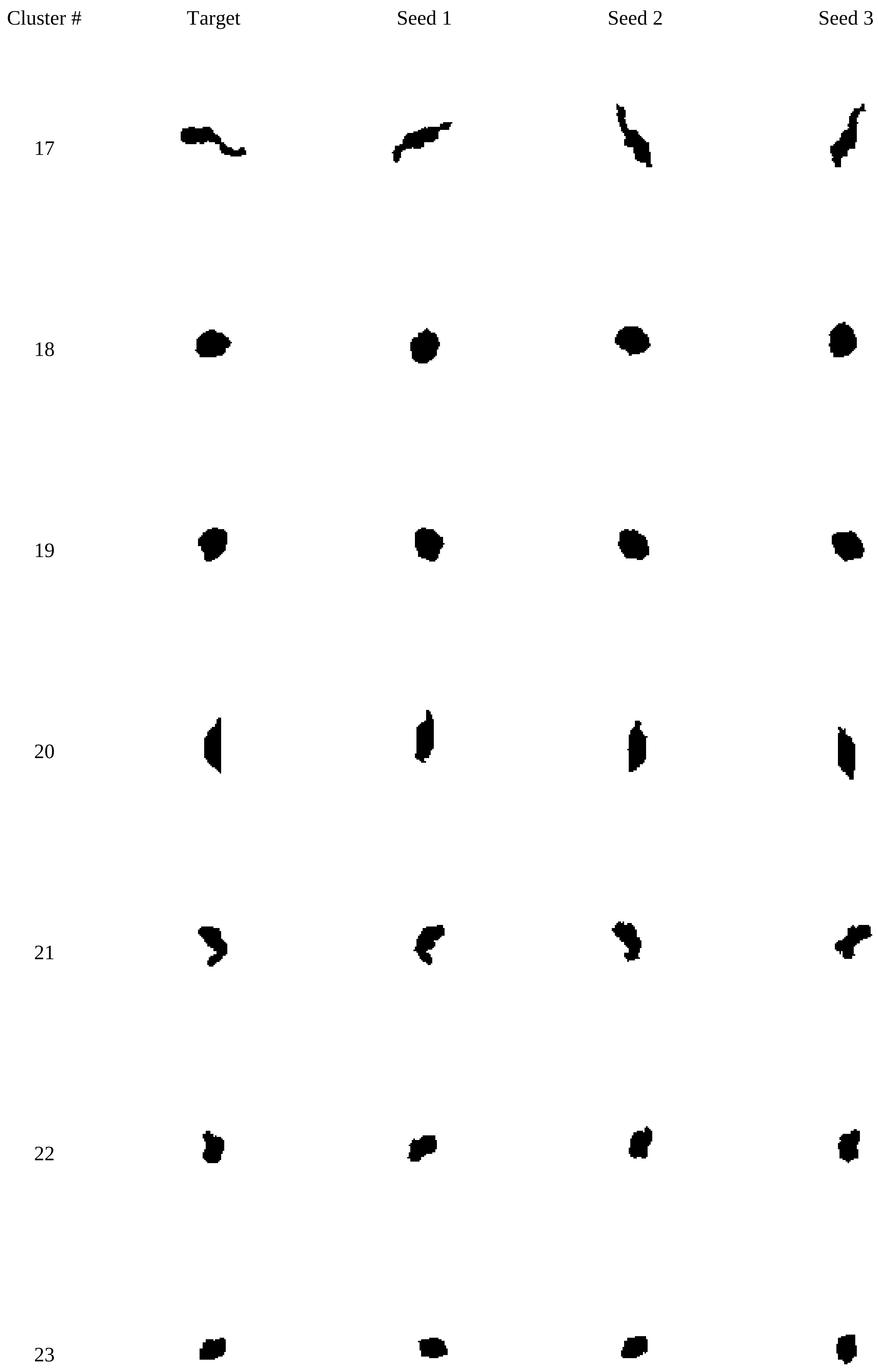

| Cluster # | Target | Seed 1 | Seed 2 | Seed 3 |
|---|---|---|---|---|
| 24 | 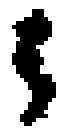 | 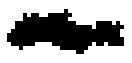 | 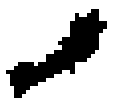 | 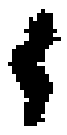 |
| 25 | 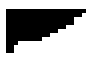 | 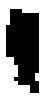 | 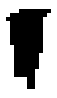 | 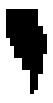 |
| 26 | 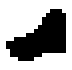 | 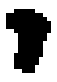 | 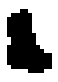 | 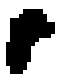 |
| 27 | 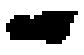 | 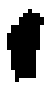 | 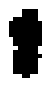 | 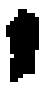 |
| 28 | 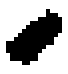 | 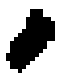 | 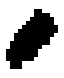 | 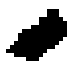 |
| 29 | 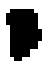 | 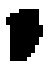 | 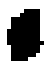 | 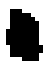 |
| 30 | 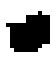 | 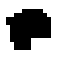 | 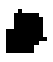 | 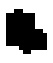 |

| Cluster # | Target | Seed 1 | Seed 2 | Seed 3 |
| --- | --- | --- | --- | --- |
| 31 | 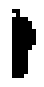 | 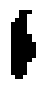 | 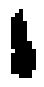 | 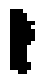 |
| 32 | 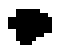 | 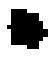 | 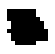 | 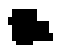 |
| 33 | 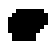 | 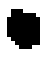 | 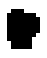 | 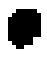 |
| 34 | 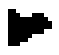 | 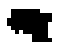 | 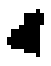 | 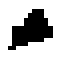 |
| 35 |  | 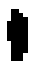 | 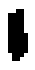 | 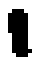 |
| 36 | 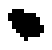 | 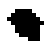 | 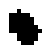 | 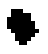 |
| 37 | 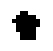 | 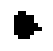 |  |  |

| Cluster # | Target | Seed 1 | Seed 2 | Seed 3 |
|---|---|---|---|---|
| 38 | | | | |
| 39 | | | | |
| 40 | | | | |
| 41 | | | | |
| 42 | | | | |